# Single-photon emission associated with double electron capture in $F^{9+}$ + C collisions

T. Elkafrawy[1,2,3,*], A. Simon[4], J. A. Tanis[1], and A. Warczak[5]

[1]*Department of Physics, Western Michigan University, Kalamazoo, Michigan 49008, USA*
[2]*Department of Physics, Faculty of Science, Ain Shams University, Abbassia, Cairo 11566, Egypt*
[3]*ASRT-ENHEP, CMS, CERN, Route de Meyrin 385, 1217 Meyrin, Geneva, Switzerland*
[4]*Department of Physics, University of Notre Dame, Notre Dame, Indiana 46556, USA*
[5]*Institute of Physics, Jagiellonian University, Krakow, Poland*
[*]Email address: tamer.elkafrawy@cern.ch

Radiative double electron capture (RDEC), the one-step process occurring in ion-atom collisions, has been investigated for bare fluorine ions colliding with carbon. RDEC is completed when two target electrons are captured to a bound state of a projectile simultaneously with the emission of a single photon. This work is a follow-up to our earlier measurement of RDEC for bare oxygen projectiles, thus providing a recipient system free of electron-related Coulomb fields in both cases and allowing for the comparison between the two collision systems as well as with available theoretical studies. The most significant mechanisms of x-ray emission that may contribute to the RDEC energy region as background processes are also addressed.

## I. INTRODUCTION

The one-step process of radiative electron capture (REC) [1,2] occurs when a loosely-bound target electron is captured into projectile and can be treated as the time-reversed process of photoionization (PI) [3]. Measurement of REC into bare ions [2] offers a clean method for exploration of photoionization of H-like ions, allowing for observation of the γ-e interaction in its purest form. Similarly, double photoionization (DPI) [4] can be studied by the investigation of the time-reversed process of radiative double electron capture (RDEC) by bare projectile ions in collisions with atoms. Accordingly, RDEC is a one-step process in ion-atom collisions occurring when two target electrons are captured to a bound state of the projectile during a single collision with the simultaneous emission of a single photon. The emitted photon has approximately double the energy of the photon emitted due to REC. RDEC was first predicted by Miraglia and Gravielle [5] and has been addressed over more than two decades theoretically [6–12] and experimentally [13–18]. Using bare ions as projectiles in the RDEC experiments allows the target electrons to be transferred without interaction with projectile electrons, enabling accurate investigation of the electron-electron interaction during the process. Thus, studying RDEC into bare projectiles provides a means to obtain a proper description of the two-electron wave function in the projectile continuum.

To optimize for the best experimental conditions under which RDEC can be observed, solid targets were chosen in nonrelativistic collisions to obtain the highest rates of double-electron capture [13]. Solid-state targets were also proposed theoretically in slow collisions with multicharged ions [7] where valence electrons behave as quasifree particles with a characteristic velocity considerably smaller than that of the projectile even for ~1 MeV/u collisions. Such comparisons, in addition to the theoretical predictions [7,8] suggesting projectiles of moderate $Z$ for lower-energy collisions, were the motivation to conduct the RDEC experiments under these conditions.

## II. KINEMATICS

### A. REC

The peaks of REC and RDEC have widths broader than the peaks of the characteristic x rays and defined by the Compton profile of the target electrons [19], which describes the momentum distribution of the bound electrons within the target atom. This momentum is the projection of the intrinsic momentum vector of the bound electron on the collision axis, defined by the direction of the projectile velocity. The Compton-profile width increases with the increase of the atomic number and the magnitude of the electron binding energy. Hence, the distribution becomes broader for heavier targets and for the capture of inner-shell electrons. REC is completed when a target electron is captured to a projectile bound state with the simultaneous emission of a photon of energy $\hbar\omega_{REC}$ as shown in Fig. 1(a). The photon energy is determined in the nonrelativistic domain [20] from the conservation of energy by

$$\hbar\omega_{REC} = K_e - B_p + B_t + \upsilon_p p_z, \qquad (1)$$

where $K_e$ is the kinetic energy of the captured target electron as calculated in the rest frame of the projectile, while $B_t$ and $B_p$ are negative values, by convention, denoting the binding energies of the target electron before and after being captured, respectively. The quantities $\upsilon_p$ and $p_z$ designate the projectile velocity, which is the same as the velocity of the captured target electron in the rest frame of projectile and the projection of the intrinsic momentum vector of the bound target electron on the collision axis Z, which is defined by the direction of the projectile velocity, respectively. REC peaks shift to lower energies as the projectile velocity decreases.

More than forty years before REC was observed [1,20], a reliable theoretical prediction of the total radiative recombination (RR) cross section was derived by Stobbe [21] and later estimated by Bethe and Salpeter (*B-S*) [22]. The *K*-REC cross section $\sigma_{REC}^{1s^1}$ will be equal to that for RR if the captured electron is loosely bound. The *B-S* formula is given for a bare nucleus and per target electron within the non-relativistic dipole approximation by

$$\sigma_{REC}^{B-S} = (9.16 \times 10^{-21}) \left(\frac{\kappa^3}{1+\kappa^2}\right)^2 \frac{\exp(-4\kappa \tan^{-1}(1/\kappa))}{1-\exp(-2\pi\kappa)}, \quad (2)$$

where $\kappa$ is the Sommerfeld parameter defined [8] by

$$\kappa = \frac{p_{ep}}{p_e} = \frac{\upsilon_{ep}}{\upsilon_p} = \frac{Z_p e^2}{\hbar \upsilon_p}. \quad (3)$$

where $p_{ep}$ is the average momentum of the target electron after being captured and $p_e$ is the momentum of the target electron while being captured, both to the projectile *K* shell in its rest frame. The adiabacity parameter $\eta$, a value that judges how fast ($\eta > 1$) or slow ($\eta < 1$) the collision is, can be defined by

$$\eta = \kappa^{-2} = \frac{K_e}{K_{ep}} = \left(\frac{\upsilon_p}{\upsilon_{ep}}\right)^2 \simeq (40.31) \frac{K_e(\text{MeV})}{Z_p^2}, \quad (4)$$

where $\upsilon_{ep}$ is the average velocity of the target electron after being transferred to the projectile *K* shell.

The angular distribution of *K*-REC as described by its differential cross section within the dipole approximation in nonrelativistic collisions is given in Ref. [23] as

$$\frac{d\sigma_{REC}^{1s^1}}{d\Omega} = \sigma_{REC}^{\exp} \left(\frac{3}{8\pi}\right) \sin^2 \theta, \quad (5)$$

where $\theta$ is the x-ray observation angle with respect to the beam direction. The corresponding predicted differential *K*-REC cross section can be obtained if $\sigma_{REC}^{\exp}$ is replaced by $N_t \sigma_{REC}^{B-S}$ where $N_t$ is the number of target electrons

## B. RDEC

During RDEC, the photon is emitted simultaneously with the capture of two bound target electrons into projectile bound states as illustrated in Fig. 1(b). The energy of the emitted photon is then given by

$$\hbar \omega'_{RDEC} = 2K_e - B_p^{(1)} - B_p^{(2)} + B_t^{(1)} + B_t^{(2)} + (\upsilon_p p_z)^{(1)} + (\upsilon_p p_z)^{(2)}, \quad (6)$$

where the indices (1) and (2) denote the first and the second captured target electrons.

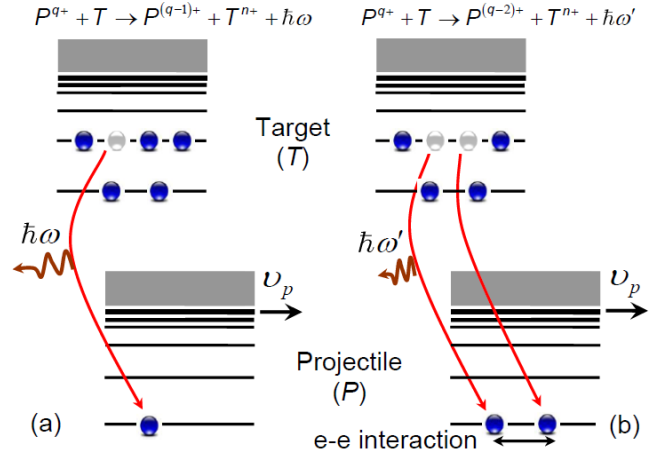

FIG. 1. Schematic diagram for (a) REC and (b) RDEC, showing single and double electron capture, respectively, both into a bare ion accompanied by photon emission.

The two captured electrons experience a mutual Coulomb repulsion, while the emitted photon satisfies the law of energy conservation between the initial and final states of the collision partners. The target electrons can be captured from the same orbit or from two different orbits to the projectile *K* shell (*KK*-RDEC) or to both the *K* and *L* shells (*KL*-RDEC). RDEC can also occur by capture of both electrons to the *L* or higher shells, but this is not seen due to the considerably lower energy of the photon emitted. In the present work the RDEC photon with the lowest energy is expelled when two *K*-shell target electrons are captured to the projectile metastable state $1s^1 2s^1$, while the photon of highest energy is released if two valence (*L*-shell) target electrons are captured to the projectile ground state $1s^2$. The angular distribution of RDEC photon emission has not been reported yet in the literature, but it is assumed to be the same as that for REC photons given by Eq. (5) [23].

Attempts to observe RDEC at the GSI accelerator facility in Darmstadt [13,14,18] did not result in the observation of RDEC. This was presumably due to the poor statistics of the data collected in the limited beam time devoted to the measurements and also to the background processes. Although RDEC was not seen, an upper limit of the *KK*-RDEC cross section during two experiments [13,14] was deduced, while the corresponding cross-section ratios of RDEC/REC and *KL*-RDEC/*KK*-RDEC were calculated based on various theoretical predictions [5–8,46]. A mid-*Z* projectile of nonrelativistic energy was used for the first attempt to observe RDEC for 11.4 MeV/u Ar$^{18+}$ + C [13] and a deduced upper limit of 5.2 mb/atom was in fair agreement with the theories from Refs. [6,7]. The relativistic heavy-ion collision system of 297 MeV/u U$^{92+}$ + Ar [14] was chosen for the second attempt to measure RDEC for which a deduced upper limit of 10 mb/atom was found to be two orders of magnitude lower than the relativistic prediction from Ref. [6] and two orders of magnitude higher than the prediction from Ref. [7]. A third unsuccessful attempt to observe RDEC was performed using the collision systems of 30 MeV/u $Cr^{24+}$ + He and $N_2$ with a detection lower limit of $10^{-4}$ mb/atom [18].

The first successful observation of RDEC was reported for the collision system of 2.38 MeV/u $O^{8+}$ + C [15]. In the present work an evidence for RDEC is reported for the collision system of 2.21 MeV/u $F^{9+}$ + C. It is noted that two preliminary publications of this work have already been presented [16,17]. In this work, a different approach of data analysis is used and corrections are made that affect the obtained cross sections, giving the best estimate of the RDEC cross sections for $F^{9+}$ + C collisions. The present measurements are compared with those of the previous experimental [15] and theoretical [6–8,12] studies. The most important background processes that can overlap with the desired processes were also investigated.

### C. Background Processes

In the present study, several background processes may contribute to the x-ray spectra and overlap with the REC and RDEC events of interest. Possible contributions from electron-nucleus (e-n) bremsstrahlung [24], nucleus-nucleus (n-n) bremsstrahlung, also called nuclear bremsstrahlung (NB) [25,26], electron-electron (e-e) bremsstrahlung [27], the two-step process of uncorrelated double radiative electron capture (DREC) [28], and pileup are considered in the analysis of the observed x rays. Considering these background processes enables extraction of the RDEC contributions more accurately from the singles x rays and ensures that their contribution to the energy domain of interest is properly estimated.

Of the background processes, e-n bremsstrahlung radiation dominates and is emitted when an electron scatters from an ion and suffers a speed reduction due to the Coulomb interaction between the incoming electron and the nucleus. By the conservation of energy a photon is emitted with an energy equal to the loss in electron kinetic energy. Assuming that all incident electrons have the same kinetic energy and due to the different impact parameters of incident electrons, not all the electrons are decelerated to the same degree, which results in a continuous range of x-ray emission. Soft x rays are emitted in the case of large impact parameters, while hard x rays correspond to small impact parameters. The captured electron in ion-atom collisions can encounter various mechanisms of e-n bremsstrahlung in the vicinity of the projectile such as radiative electron capture to continuum (RECC) [29], radiative ionization (RI) [30,31], and secondary electron bremsstrahlung (SEB) [32].

RECC can be treated as quasi-free electron bremsstrahlung (QFEB) [33] and was first observed by Kienle et al. [20]. During RI, the target electron is not captured to the projectile continuum (QFEB) but ionized away from the projectile with simultaneous emission of a photon. If the target electron is ionized and transferred to the projectile continuum then the process is QFEB as a special case of RI. SEB is a two-step process and occurs if a target electron is ejected by projectile impact and then scattered by the Coulomb field of another target nucleus. However, SEB was found to be less important for low-Z targets such as Be and C [34] based on the cross section calculated from the Koch bremsstrahlung formula [24].

The n-n bremsstrahlung was treated theoretically [25] and first observed for heavy-ion collisions [26], provided that this emission component is isolated from the other x-ray emission processes. The e-e bremsstrahlung originates from the interaction between the incoming electrons and the bound target electrons. Its contribution was found to be $1/Z$ of the net bremsstrahlung radiation [35] based on which it can be neglected in the case of high-Z targets. In contrast, the cross section of e-n bremsstrahlung scales closely to $Z^2$ for unshielded nuclei, while there is no simple Z-dependence for shielded nuclei [36].

DREC can contribute to the REC energy domain in the spectrum of x rays associated with double capture when two REC photons are emitted due to the capture of two uncorrelated target electrons into the projectile in a single collision. Due to its exceedingly small probability [28], DREC has a negligible contribution to the REC events accompanied by nonradiative electron capture (NRC) [37]. It is also highly unlikely that DREC contributes to the RDEC energy domain in the double-capture channel when the two DREC photons are emitted in the same direction and registered as a single photon of double energy.

In the case of high-beam intensities, the rate of the collisions increases, allowing for an increase in the pileup probability, i.e., the probability to have two ions emitting two photons detected simultaneously and registered as a single photon of double the energy within the RDEC energy domain in the single-capture channel. While pileup as a mechanism applies to any source of x-ray emission, pileup of interest in the present case is that originating from the superposition of two REC photons.

### III. EXPERIMENT

This work was performed using the tandem Van de Graaff accelerator facility at Western Michigan University. A bare fluorine beam with energy of 2.21 MeV/u was obtained following the production of negative fluorine ions from a source of negative ions by cesium sputtering (SNICS II) and the subsequent acceleration. In addition, a beam of 3 MeV $H^+$ from the same ion source was used to conduct elemental analysis of the carbon targets utilizing proton-induced x-ray emission (PIXE) measurements.

A 90° analyzing magnet following the accelerator was used to select the desired charge state and energy. For the fluorine ions ($F^{7+}$), a carbon-foil post-stripper followed the analyzing magnet to produce higher charge states ($F^{8+}$, $F^{9+}$) that cannot be reached using only the gas stripper at the accelerator terminal. The energy of the $F^{7+}$ beam emerging from the analyzing magnet was slightly reduced after passing through a post stripper that is 10.0(1.5) $\mu g/cm^2$ thick by about 0.3% of the incoming beam energy for the same charge state [38]. A switching magnet then directed the appropriate charge state into the beam line.

The experiment was set up as shown in Fig. 2, which is similar to the one used for the $O^{8+}$ + C experiment [15]. A holder with an Al frame carrying a target carbon foil of mass areal density of 10.9(1.6) $\mu g/cm^2$ was mounted and positioned at 45° to the incoming beam, corresponding to an atomic target thickness of $7.7(1.1) \times 10^{17}$ atom/cm². X-ray attenuation due to absorption does not appreciably occur for a beam passing at 45° inclination through the thin carbon foil based on attenuation coefficients of 0.33–0.04% for the x-ray energy range of 2–4 keV, respectively [39]. An empty frame was used in one of the target positions so that the background could be determined and to ensure that the emitted x rays originate only from the collisions with the carbon foils and not, for instance, from the aluminum frame or any other impurities that might exist on the frame surface. A 2-mm-wide collimator was used to ensure good beam collimation at the target as illustrated in Fig. 2.

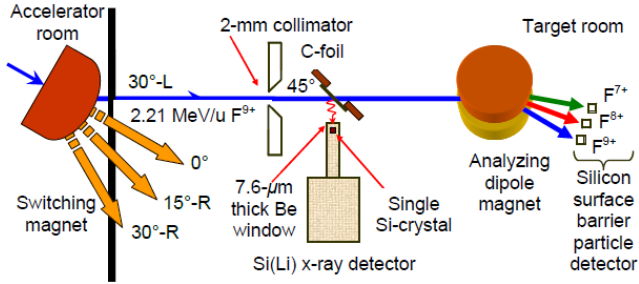

FIG. 2. Schematic diagram of the experiment setup in the target room in a top view of random scale (not 1:1).

A Si(Li) detector positioned at 90° to the incident beam was used to detect the emitted x rays. The detector had a Be window of thickness 7.6(1.1) $\mu$m and a single Si(Li) crystal of active diameter, thickness, and active area of 6 mm, 5 mm, and 28 mm², respectively. From the center of the C foil, the crystal was at a distance of 19.0(0.76) mm including 5.0(0.6) mm behind the Be window. The foil was mounted at the center of the beam line, giving a detection solid angle $\Delta\Omega$ of 0.078(0.003) sr. The dimensions given for the Be window and Si crystal correspond to a detection efficiency of 75–100% in the x-ray energy domain of 1.5–15 keV, respectively. The Si(Li) detector used had an actual energy resolution (FWHM) of 240 eV at the energy of the Mn $K\alpha$ characteristic line (~5.9 keV) obtained from a standard $^{55}$Fe radioactive source.

Charge-changed projectile ions with charge states $q$-2, $q$-1, as well as the charge state $q$ of the primary beam were separated by a dipole analyzing magnet and registered individually by three ion-implanted silicon surface-barrier particle detectors. These charge states occurred in the approximate count ratios of 1.0:18.3:9.2 for $q$-2, $q$-1, and $q$, respectively. Emitted photons were analyzed in coincidence with ions of outgoing projectile charge states $q$-2, $q$-1, and $q$, with the data acquisition system providing the required coincidence techniques to isolate the correlated processes.

## IV. RESULTS AND DISCUSSION

### A. Singles x rays and coincidence spectra

PIXE analysis of the carbon foil target was performed with 3-MeV protons to determine if there is a contribution of x rays from impurities in the foil to x rays in the energy domain of interest (up to 5 keV), with the line intensities giving estimates of the impurities percentages in the target.

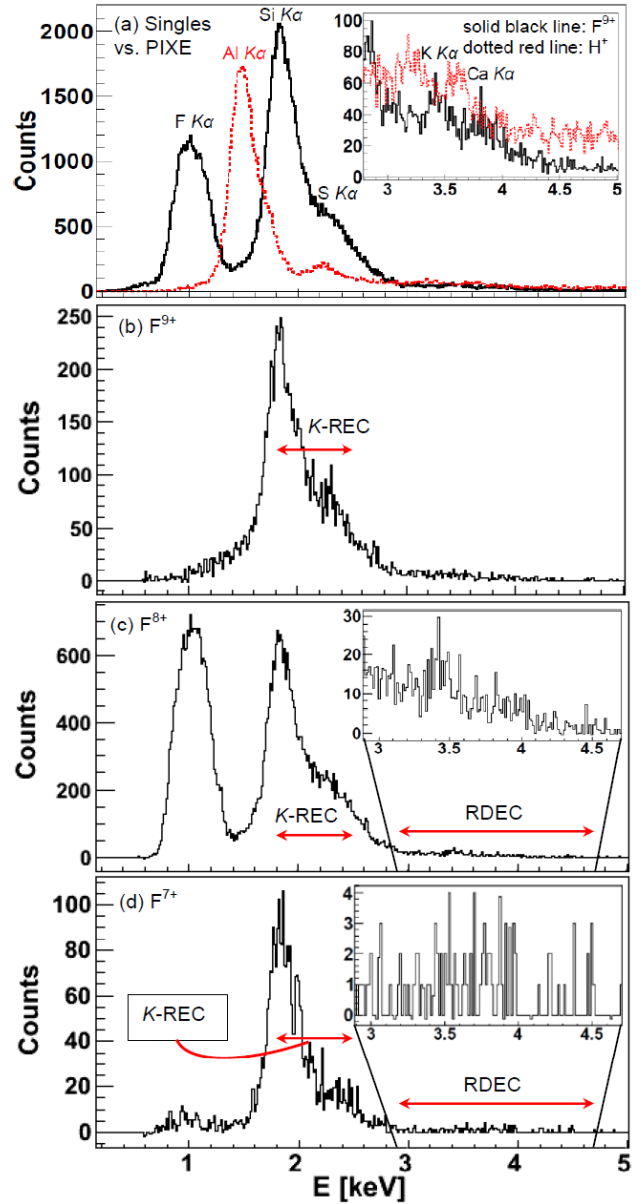

FIG. 3. Measurements for (a) singles x rays (in solid black) with PIXE measurements (in dotted red) superimposed using 3-MeV protons as well as those for x rays with randoms subtracted associated with (b) $F^{9+}$, (c) $F^{8+}$, and (d) $F^{7+}$, all for 2.21 MeV/u $F^{9+}$ on the same C foil.

In Fig. 3(a), the PIXE results are compared with the spectrum of singles x rays. PIXE analysis of the target carbon foil showed evidence for contaminations with Si, S, K, and Ca, while an Al $K\alpha$ line was measured only in the case of PIXE. The carbon foil was probably contaminated during its floating process, while the Al $K\alpha$ line is likely caused by the larger scattering angles of the protons than the $F^{9+}$ ions (this was calculated to be a ratio of ~1.6) or by a different steering of the proton beam. This will happen if the protons hit the aluminum frame on which the target carbon foil is mounted as a result of touching the edge of the aperture just prior to the foil.

The characteristic $K\alpha$ lines of all contaminations come at lower energies in the case of PIXE than that in the case of the $F^{9+}$ beam, caused by the shift due to multiple ionization of the target by incident $F^{9+}$. The shifts are found to be in the range of 15–25 eV per $L$-shell vacancy [40], which agrees with measured shifts for the lines Si $K\alpha$ (~60 eV) up to Ca $K\alpha$ (~100 eV). X rays resulting from $F^{9+}$ + C collisions were measured in coincidence with no-charge change ($F^{9+}$), single capture ($F^{8+}$), and double capture ($F^{7+}$) and are shown in Figs. 3(b,c,d), respectively, where random x rays were subtracted from each of the spectra.

The F $K\alpha$ line is clearly observed in the spectrum of x rays associated with $F^{8+}$, while this line is seen in the $F^{7+}$ channel with intensity that is 1% of that observed in the $F^{8+}$ channel, and essentially no evidence is seen for this line in the x-ray spectrum associated with $F^{9+}$. Similar results were found for the $q$-1 and $q$-2 channels for $O^{8+}$ + C collisions [15] (the $q$ channel was not registered in this work). This shows that there is essentially no crossover of the characteristic x rays from the F $K\alpha$ line among the various spectra. Contaminant x rays from the target, however, are expected to contribute to all the channels shown in Fig. 3. The Si and S $K\alpha$ lines overlap with the $K$-REC structure, while the low-energy (2.9–4.0 keV) RDEC events overlap with the K and Ca $K\alpha$ lines as shown in Fig. 3(a). Calculated energies with uncertainties of the REC and RDEC lines are listed in Table I for the current experiment.

TABLE I. Calculated energies (in eV) given in ascending order of the REC and RDEC lines for 2.21 MeV/u $F^{9+}$ + C.

| X-ray emission line | E (eV) |
| --- | --- |
| $K$-REC of $K$-shell electron | 2018(16) |
| $K$-REC of valence electron | 2306(18) |
| $KL$-RDEC of two $K$-shell electrons | 3172(25) |
| $KL$-RDEC of valence and $K$-shell electrons | 3456(28) |
| $KL$-RDEC of two valence electrons | 3740(30) |
| $KK$-RDEC of two $K$-shell electrons | 3893(31) |
| $KK$-RDEC of valence and $K$-shell electrons | 4177(33) |
| $KK$-RDEC of two valence electrons | 4461(36) |

### B. REC calculation

The contamination lines observed by means of PIXE analysis in the REC energy range (Si $K\alpha$ and S $K\alpha$) are found to be distributed between the $F^{7+}$, $F^{8+}$, and $F^{9+}$ channels. REC appears in the $F^{8+}$ channel as the normal channel and also appears in the $F^{7+}$ channel if two target electrons are captured independently to the same projectile, one radiatively (REC) and the other nonradiatively (NRC). Thus, the estimation for the probability of having NRC [37] accompanying REC [22] helps avoid underestimating the REC cross section by considering the additional events that appear in the double-capture channel. The probability of having NRC and REC accompanying each other was found to be $2.1 \times 10^{-5}$ based on a $K$-REC probability of $4.1 \times 10^{-4}$ and under the experimental conditions of the current work. It is also likely that some of the $K$-REC events will appear in the $F^{9+}$ channel, which can occur from re-stripping of the $F^{8+}$ component of beam in the target foil promptly after $K$-REC is completed. However, this $K$-shell ionization has a probability of about one order of magnitude less than that of $L$-shell ionization [41]. Since the energy domain of $K$-REC was dominated by the Si and S $K\alpha$ lines, taking into account shifts of ~60 eV of each of the two lines due to multiple ionization, the $K$-REC structure cannot be extracted reliably from the data. Instead, the total $K$-REC cross section was calculated from the Bethe-Salpeter formula [22] as a reliable source for this purpose and found to be 525 b/atom.

### C. Analysis of background processes

The collision energy of 2.21 MeV/u for the current experiment corresponds to a kinetic energy of 1.2 keV for a captured target valence electron in the projectile frame. The x-ray contributions from QFEB, SEB, and NB were studied for 2-MeV protons incident on carbon [31], which is about the same as the present collision system. This, along with calculations from Ref. [42], show that QFEB and SEB are characterized by the maximum emitted photon energy $T_r$ (found to be 1.2 keV) and the maximum energy transferred from a projectile to a free electron at rest $T_m$ (found to be 4.8 keV) and are predominant in the photon-energy ranges $\hbar\omega \leq T_r$ and $T_r < \hbar\omega < T_m$, respectively. This implies that the RDEC energy range of 2.9–4.7 keV does not overlap with QFEB ($\leq 1.2$ keV), while SEB was found to give a negligible contribution to x rays in the energy range of 1.2–4.8 keV for low-Z targets [34], such as the carbon target foil used for the this work. RI is of more significance in slow than in fast collisions for which it is hard to derive a scaling rule [30]. A study [31] shows that NB has a very small contribution to the continuous x-ray spectrum compared to the e-n bremsstrahlung processes for x-ray energies $\leq 10$ keV. Another study for the collision system of 1.5 MeV $H^+$ + Al [43] shows that NB has a minimum contribution when the x rays are measured at 90°, such as the case of this work, and in specific, it is concentrated at angles within 10° to the line of the colliding beams [44]. The e-e bremsstrahlung is not taken into account in most of the bremsstrahlung measurements due to its small contribution to the total bremsstrahlung emission [35].

The $KK$-DREC probability for this work is found to be $2.4 \times 10^{-9}$ based on the cross section given in Table II, as obtained from Ref. [28], which is five and three orders of magnitude lower than values of the calculated $K$-REC and the total $KK$-RDEC cross sections, respectively. This represents a negligible contribution of DREC to the REC and RDEC energy ranges. Considering pileup, the counting rate of photons encountered in the present work was ~4 counts/s as a result of the very low beam intensity used (<1 pA). Based on the total $K$-REC cross section of 525 b/atom [22] and the given target thickness, the probability of pileup from REC photons was found to be $1.3 \times 10^{-12}$, which, compared to the total RDEC probability of $7.1(3.6) \times 10^{-6}$ (based on the total RDEC cross section given in Table II), implies a negligible contribution to the RDEC photons in the single-capture channel and in turn in the double-capture channel when REC accompanies NRC.

### D. RDEC analysis

The double-capture channel ($F^{7+}$) is the proper channel for the RDEC events to appear. However, following formation of the $F^{7+}$ beam component, one of the two captured electrons can be promptly re-stripped in the remaining target that the ion passes through, causing the RDEC event to appear in the single-capture channel ($F^{8+}$). The probability for this stripping when a $K$-shell electron is ejected, i.e., $F^{7+} \rightarrow F^{8+}$, following a $KK$-RDEC event was calculated to be ~0.39 [45] based on an estimated cross section of 1.0 Mb and the target thickness of the C foil used for the current measurement. For an $L$-shell electron it is certain (>100%) [45] that it will be re-stripped from the $F^{7+}$ ions associated with $KL$-RDEC.

The x rays associated with $F^{7+}$ and $F^{8+}$ over the entire RDEC energy domain originate from $KL$-RDEC and $KK$-RDEC as well as a small contribution from the high-energy REC tail and contamination lines due to the K and Ca $K\alpha$ lines where both contributions overlap mostly with $KL$-RDEC and partially with $KK$-RDEC as shown in Fig. 3(a). In order to correct the x-ray spectra associated with $F^{7+}$ and $F^{8+}$ for these contributions, the $K$-REC high-energy tail was fitted to the $F^{7+}$ and $F^{8+}$ spectra as shown in Figs. 4(a,b), and subtracted from each of the two spectra. The singles x rays over the RDEC energy range are largely due to the contamination lines from K and Ca (as determined from the PIXE analysis) as seen in Fig. 3(a). The REC high-energy tail was also subtracted from this spectrum, and then the subtracted singles x rays were normalized to an intensity where contamination lines in the singles x rays match the same contamination lines in the x-ray spectra associated with $F^{7+}$ and $F^{8+}$ as shown in Figs. 4(c,d). The normalized singles x-ray spectra were subtracted from the $F^{7+}$ and $F^{8+}$ spectra as a means to subtract the contributions of the two contaminant lines due to K and Ca from the $F^{7+}$ and $F^{8+}$ spectra. The leftover events from Figs. 4(c,d) were added together to give the total events due to RDEC in the energy range of 2.9–4.7 keV as seen in Fig. 4(e). The $KL$-RDEC (2.9–4.0 keV) and $KK$-RDEC (3.6–4.7 keV) regions as well as the expected RDEC lines are also indicated in the same figure. This is where the present analysis differs significantly from our earlier analysis [16,17], and should give more accurate values for the measured RDEC cross sections.

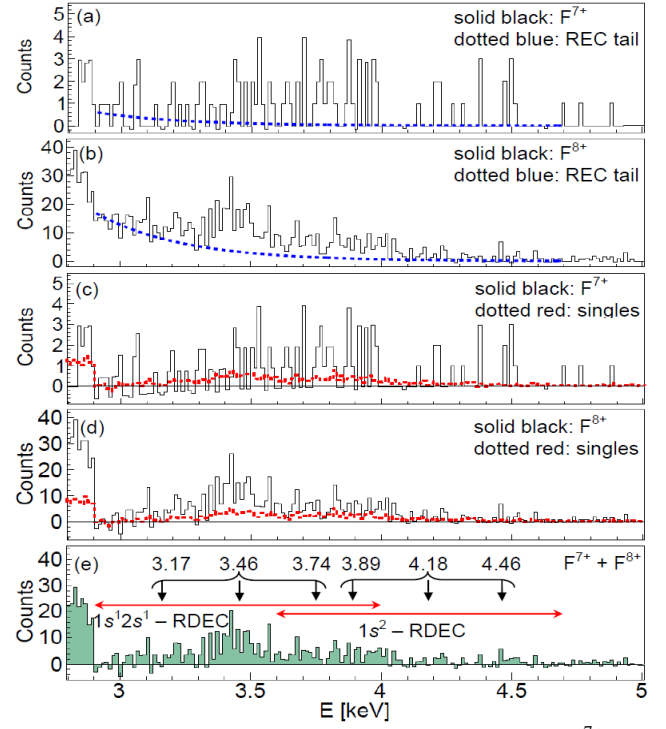

FIG. 4. Spectra showing x rays associated with (a) $F^{7+}$ and (b) $F^{8+}$, both with the REC high-energy fitting superimposed, (c) $F^{7+}$ and (d) $F^{8+}$, both with singles x rays normalized to, and superimposed on, each of them, and with the REC high-energy tail subtracted from the spectra, and (e) sum of the $F^{7+}$ and $F^{8+}$ x-ray spectra with the normalized singles x rays subtracted from each of them, giving the total RDEC structure (the calculated energies of RDEC lines are indicated in keV for 2.21 MeV/u $F^{9+}$ + C).

### E. Measured RDEC cross sections

The measured differential $KL$-RDEC and $KK$-RDEC cross sections at 90° for the current collision system are found to be 0.54(0.22) and 0.55(0.33) b/sr.atom, respectively, giving a total differential RDEC cross section of 1.09(0.55) b/sr.atom. The corresponding total $KL$-, $KK$-, and RDEC cross sections are given in Tables II and III based on Eq. (5). For comparison, the measured total RDEC cross sections [15] and the obtained upper limits of the $KK$-RDEC cross sections [13,14] for the previous experiments are given in Table III. A very good consistency between the current results and those obtained for the $O^{8+}$ + C experiment [15] was achieved, which is expected for the two systems having very similar kinematic conditions. Alongside with the experimental data, the corresponding calculated values of *KL*- and *KK*-RDEC cross sections from the available theories [6,7,12,46] are listed in Table III, which are seen to be up to four orders of magnitude smaller than the experimental values. The closest theoretical prediction of the *KK*-RDEC cross section to our measured value is obtained from the first approximation introduced in Ref. [12] where the electrons are assumed to be distributed homogeneously in the entire volume of the atom (measured value is underestimated by a factor of ~5).

TABLE II. Cross sections in descending order for REC, RDEC, and the other background processes for 2.21 MeV/u $F^{9+}$ + C.

| Atomic process | Measured or calculated | Energy domain (keV) | Cross section (b/atom) |
|---|---|---|---|
| *K*-REC ($1s^1$) [22] | Calculated | 1.8–2.5 | $5.3 \times 10^{+2}$ |
| *K*-NRC/*K*-REC ($1s^2$) [37] | Calculated | 1.8–2.5 | $2.7 \times 10^{+1}$ |
| Total RDEC ($1s^2 + 1s^1 2s^1$) | Measured | 2.9–4.7 | 9.1(4.6) |
| *KL*-RDEC ($1s^1 2s^1$) | Measured | 2.9–4.0 | 4.5(1.8) |
| *KK*-RDEC ($1s^2$) | Measured | 3.6–4.7 | 4.6(2.8) |
| *KK*-DREC ($1s^2$) [28] | Calculated | 1.8–2.5 | $3.1 \times 10^{-3}$ |
| Pileup ($1s^1 + 1s^1$) [22] | Calculated | 3.6–5.0 | $1.7 \times 10^{-6}$ |

TABLE III. Measured versus calculated *KL*-RDEC and *KK*-RDEC cross sections for the experiments that gave results. The abbreviations Mis (Che previously), Nef, Mik, Yak, PDB, and Exp stand for Mistonova (Chernovskaya previously), Nefiodov, Mikhailov, Yakhontov, principle of detailed balance, and experiment, respectively.

| $Z_p$ | $E_p$ (MeV/u) | $\kappa$ | $Z_t$ | $\sigma_{RDEC}^{1s^1 2s^1}$ (b/atom) | | | $\sigma_{RDEC}^{1s^2}$ (b/atom) | | | | |
|---|---|---|---|---|---|---|---|---|---|---|---|
| | | | | Che [11] | Nef [8] | Exp | Mis [12] | Mik [7] | Yak [6] | PDB [46] | Exp |
| 18 [13] | 11.4 | 0.84 | 6 | --- --- | 2.2E-3 | --- | 0.12[a] 4.3E-3[b] | 3.5E-3[c] 2.9E-6[d] | 1.9E-3 | 0.045 | ≤5.2E-3 |
| 92 [14] | 297 | 0.84 | 18 | --- --- | 1.7E-5 | --- | 1.7E-3 3.1E-6 | 2.7E-5 8.4E-10 | 5.0 | 5.8E-3 | ≤0.010 |
| 8 [15] | 2.38 | 0.82 | 6 | 0.050[a] 2.0E-3[b] | 0.10 | 2.3(1.3) | 0.55 0.019 | 0.14 1.2E-4 | 0.14 | 0.23 | 3.2(1.9) |
| 9 [16] | 2.21 | 0.96 | 6 | --- --- | 0.24 | 4.5(1.8) | 0.94 0.033 | 0.27 2.2E-4 | 0.12 | 0.18 | 4.6(2.8) |

[a]First approximation involving the entire atom
[b]Second approximation involving only the *K* shell
[c]For the capture of two *K*-shell target electrons
[d]For the capture of two valence target electrons

The measured *KL*-RDEC/*KK*-RDEC cross-section ratio $R'' = \sigma_{RDEC}^{1s^1 2s^1} / \sigma_{RDEC}^{1s^2}$ for this work was found to be a factor of 1.3 greater than the measured value in the case of the $O^{8+}$ experiment [15]. Nefiodov *et al.* [8] predicted that the ratio $R''$ is enhanced drastically for slow collisions over the range $1 < \kappa < 10$, which led to the prediction of the *KL*-RDEC cross section. This, along with the prediction from Ref. [7] of the *KK*-RDEC cross section, was used to predict an empirical formula for $R''$ [8]. Thus, the theory is expected to work well in the domain $1 < \kappa < 10$, where $\kappa = 0.96$ for this work. The calculated values of $R''$ based on the only available theory [8] to predict this ratio are listed in Table IV compared to the RDEC experiments [15,16] that gave results.

TABLE IV. Measured versus calculated values of $R'' = \sigma_{RDEC}^{1s^1 2s^1} / \sigma_{RDEC}^{1s^2}$ for the RDEC experiments performed to date, ordered chronologically. The abbreviations Nef and Exp stand for Nefiodov and experiment, respectively.

| $Z_p$ | $E_p$ (MeV/u) | $\kappa$ | $Z_t$ | $R''$ | |
|---|---|---|---|---|---|
| | | | | Nef [8] | Exp |
| 18 [13] | 11.4 | 0.84 | 6 | 0.63 | --- |
| 92 [14] | 297 | 0.84 | 18 | 0.63 | --- |
| 8 [15] | 2.38 | 0.82 | 6 | 0.70 | 0.72(0.59) |
| 9 [16] | 2.21 | 0.96 | 6 | 0.90 | 0.96(0.70) |
| 24 [18] | 30 | 0.69 | 2 | 0.57 | --- |
| 24 [18] | 30 | 0.69 | 7 | 0.57 | --- |

## V. CONCLUSIONS

In the current work, the process of RDEC was measured for collisions of bare fluorine ions with a thin carbon foil at a collision energy of 2.21 MeV/u and the results are compared with our previous work for collisions of bare oxygen ions with a similar target at a collision energy of 2.38 MeV/u. PIXE analysis for the present work showed evidence for contamination of the target with Si, S, K, and Ca elements. Here, the S $K\alpha$ line, shifted due to multiple ionization, overlaps with the $K$-REC structure, while the shifted K and Ca $K\alpha$ lines cover partially the $KL$-RDEC structure. Contributions from the background mechanisms including pileup and DREC were found to be negligible.

The measured $KL$-RDEC and $KK$-RDEC cross sections for the present work showed consistency with our first observation of RDEC for $O^{8+}$ ions [15] and were underestimated significantly by the various theories [6,7,12,46] as listed in Table III. The measured $KL$-RDEC/$KK$-RDEC ratio ($R''$) was found to be consistent with the measured value in the case of the $O^{8+}$ ions [15], and both values are in agreement with Ref. [8] as the only available theory to predict this ratio as given in Table IV.

## ACKNOWLEDGEMENTS

The authors would like to thank Dr. Asghar Kayani for his help with the operation of the Van de Graaff accelerator facility at Western Michigan University.